\documentclass[reprint,superscriptaddress,amsmath,amssymb,aps,prX,]{revtex4-2}

\usepackage{dcolumn}
\usepackage{bm}
\usepackage{amssymb}
\usepackage{amsmath}
\usepackage{graphicx}
\usepackage{natbib}
\usepackage{hyperref}

\newcommand{\aap}{Astron.~Astro.}
\newcommand{\ssr}{Space Sci.~Rev.}
\newcommand{\jgr}{J.~Geophys.~Res.}
\renewcommand{\apj}{Astrophys.~J.}
\newcommand{\jgrsp}{J.~Geophys.~Res.: Space Phys.}
\newcommand{\apjs}{Astrophys.~J.~Supp.}
\renewcommand{\prl}{Phys.~Rev.~Lett.}
\newcommand{\astrnat}{Nat.~Astron.}
\newcommand{\grl}{Geophys.~Res.~Lett.}
\newcommand{\angeo}{Ann.~Geophys.}
\newcommand{\lrsp}{Living Rev.~Solar Phys.}
\newcommand{\apjl}{Astrophys.~J.~Lett.}
\renewcommand{\jpp}{J.~Plasma Phys.}
\newcommand{\araa}{Ann.~Rev.~Astron.~Astro.}
\newcommand{\rspa}{Proc.~R.~Soc.~Lond.}

\begin{document}

\title{Evidence for the helicity barrier from measurements of the turbulence transition range in the solar wind}

\author{J. R. McIntyre}
\affiliation{Department of Physics and Astronomy, Queen Mary University of London, London, E1 4NS, UK}

\author{C. H. K. Chen}
\affiliation{Department of Physics and Astronomy, Queen Mary University of London, London, E1 4NS, UK}

\author{J. Squire}
\affiliation{Department of Physics, University of Otago, 730 Cumberland St., Dunedin 9016, New Zealand}

\author{R. Meyrand}
\affiliation{Department of Physics, University of Otago, 730 Cumberland St., Dunedin 9016, New Zealand}

\author{P. A. Simon}
\affiliation{Department of Physics and Astronomy, Queen Mary University of London, London, E1 4NS, UK}

\begin{abstract}
The means by which the turbulent cascade of energy is dissipated in the solar wind, and in other astrophysical systems, is a major open question. It has recently been proposed that a barrier to the transfer of energy can develop at small scales, which can enable heating through ion-cyclotron resonance, under conditions applicable to regions of the solar wind. Such a scenario fundamentally diverges from the standard picture of turbulence, where the energy cascade proceeds unimpeded until it is dissipated. Here, using data from NASA's Parker Solar Probe, we find that the shape of the magnetic energy spectrum around the ion gyroradius varies with solar wind parameters in a manner consistent with the presence of such a barrier. This allows us to identify critical values of some of the parameters necessary for the barrier to form; we show that the barrier appears fully developed for ion plasma beta of below $\simeq0.5$ and becomes increasingly prominent with imbalance for normalised cross helicity values greater than $\simeq0.4$. As these conditions are frequently met in the solar wind, particularly close to the Sun, our results suggest that the barrier is likely playing a significant role in turbulent dissipation in the solar wind and so is an important mechanism in explaining its heating and acceleration.
\end{abstract}

\maketitle

\section{Introduction}
The dissipation of a turbulent cascade is thought to be important for the dynamics of the solar wind and other astrophysical systems \cite{2004ARA&A..42..275S, 2014Natur.515...85Z}. In the case of the solar wind it is proposed that it is responsible for its heating and acceleration \cite{1968ApJ...153..371C, 1971A&A....13..380A, 1971ApJ...168..509B, 2013LRSP...10....2B}. However, the means by which this heating takes place remains a major open question. Unlike hydrodynamic turbulence, where the dissipation occurs due to viscosity, a range of processes can be involved in plasmas. Mechanisms suggested to be relevant for the solar wind include ion-cyclotron heating \cite{1982JGR....87.5030M, 1999ApJ...518..937C, 2013PhRvL.110i1102K}, stochastic heating \cite{2010ApJ...720..503C,2017ApJ...850L..11V}, Landau damping \cite{1985A&A...142..404D, 1999JGR...10422331L, 2008JGRA..113.5103H} and magnetic reconnection \cite{2007NatPh...3..236R, 2017JPlPh..83f9009M}.  The first of these, heating through the ion-cyclotron resonance (along with stochastic heating), has the advantage of being able to account for the ion temperature being greater than the electron temperature in the solar wind \cite{2009ApJ...702.1604C} and for ions being heated preferentially in the direction perpendicular to the background magnetic field \cite{1999ApJ...518..937C}. Further, circularly polarised waves, which include ion-cyclotron waves (ICWs), are regularly observed in the near Sun solar wind \cite{2019Natur.576..237B, 2020ApJS..246...66B}. However, it was for a long time not apparent how energy could be transferred to small enough scales parallel to the background magnetic field to enable this mechanism to operate, as is required for ICWs  \cite{2009ApJS..182..310S}. Meyrand et al.\ \cite{Meyrand_2021} proposed a novel ``helicity barrier" mechanism which may enable such scales to be reached \cite{Squire_2022}. In this paper we use in situ solar wind data to investigate the presence of such a mechanism, and the conditions under which it occurs.

At scales large compared to ion scales, solar wind turbulence is often considered using reduced magnetohydrodynamics (RMHD), a fluid description of plasmas valid only at scales above the ion gyroradius, $\rho_{\mathrm{i}}$ \cite{2009ApJS..182..310S}. Turbulence in RMHD can be thought of in terms of the interaction of the counterpropagating Elsasser fields, $ \delta \boldsymbol{z}^{\pm} = \delta \boldsymbol{v} \pm \delta \boldsymbol{b}$, where $\delta \boldsymbol{v}$ is the perturbation to the velocity field and $\delta \boldsymbol{b}$ is the perturbation to the magnetic field in velocity units \cite{1950PhRv...79..183E}.  The imbalance in the energy of the two Elsasser fields, known as the cross helicity, is a conserved quantity in RMHD, causing significant differences between imbalanced turbulence and balanced turbulence. The solar wind is often highly imbalanced, particularly close to the Sun \cite{2007AnGeo..25.1913B, 2020ApJS..246...53C}, making understanding this regime important for explaining its dynamics.

The helicity barrier is an example of a mechanism only possible in imbalanced turbulence. Meyrand et al.\ \cite{Meyrand_2021} argue for the barrier using the more general finite Larmor radius magnetohydrodynamics (FLR-MHD), a fluid description of a plasma also valid at scales below the ion gyroradius that reduces to RMHD at scales large compared to $\rho_{\mathrm{i}}$ and to electron reduced MHD (ERMHD) \cite{2009ApJS..182..310S} at scales small compared to $\rho_{\mathrm{i}}$.  In contrast to RMHD, in FLR-MHD there is a conserved ``generalised helicity". At scales larger than $\rho_{\mathrm{i}}$ it reduces to the RMHD cross helicity and undergoes a forward cascade, whereas at scales smaller than $\rho_{\mathrm{i}}$ it reduces to the magnetic helicity and undergoes a reverse cascade \cite{2011PhRvL.106s1104C}. Meyrand et al. propose that these oppositely directed cascades create a barrier to the generalised helicity at the ion gyroradius, only allowing the balanced component of the energy cascade to reach smaller scales. This limits the cascade rate of the dominant Elsasser field to match that of the other Elsasser field at electron scales. Squire et al.\ \cite{Squire_2022} argue that the build up of energy at ion scales, as a result of the barrier to the cascade, may allow smaller parallel scales to be reached through critical balance (where the timescale for the propagation of Alfv\'{e}n waves equals that of the nonlinear turbulent timescale) than would otherwise occur if energy were more readily dissipated. This can then enable ion-cyclotron heating.

The lack of a constant flux when the helicity barrier is active, a fundamental difference from the standard picture of turbulence, would be expected to alter the shape of the energy spectrum. This may help explain the steep transition range in the magnetic spectrum between inertial and kinetic scales that is regularly observed in the solar wind \cite{2009PhRvL.103g5006K, PhysRevLett.105.131101, 2020PhRvL.125b5102B, 2021ApJ...915L...8D}. Meyrand et al.\ \cite{Meyrand_2021} were able to produce a transition range in simulations for the first time. However, there are other suggested mechanisms for its formation, including Landau damping of kinetic waves \cite{2008JGRA..113.5103H}, ion-scale coherent structures \cite{2016ApJ...824...47L} or increased spectral transfer rates \cite{2020PhRvL.125b5102B}.

If the helicity barrier does play a role in the formation of the transition range then the properties of that part of the spectrum should vary between regions of the solar wind where the barrier is expected to form and where it is not. Solar wind parameters which can be used to differentiate such regions include:

\begin{itemize}
\item The ion plasma beta, $\beta_{\mathrm{i}}$. The helicity barrier can only operate at $\beta_{\mathrm{i}}<1$, as the barrier results from the conservation laws of FLR-MHD, which is derived assuming $\beta_{\mathrm{i}} \ll 1$. At $\beta_{\mathrm{i}} \sim 1$ resonant interactions of fluctuations and ions can break these conservation laws, an effect which, if strong enough, will prevent the barrier from forming. 
\item The normalised cross helicity, $\sigma_{\mathrm{c}}$. Since the helicity barrier effect is caused by the inability of the (generalised) helicity to cascade below $\rho_{\mathrm{i}}$, if there is no flux of $\sigma_{\mathrm{c}}$ to small scales turbulence can dissipate normally. This predicts a critical normalised helicity flux above which a barrier forms. Assuming that $\sigma_{\mathrm{c}}$ is related to the normalised flux, we only expect the barrier to form at larger $\sigma_{\mathrm{c}}$. Consistent with this, the transition range has previously been observed to be steeper for regions of greater cross helicity in the solar wind \cite{Huang_2021, Zhao_2022, 2024NatAs.tmp...24B}. Meyrand et al.\ \cite{Meyrand_2021} and Squire et al. \ \cite{2023ApJ...957L..30S} also make an empirical prediction, based on simulations, that the perpendicular wavenumber of the break point between the inertial and transition range, $k_{\perp,1}$, varies with $\sigma_{\mathrm{c}}$ as $k_{\perp,1} \rho_{\mathrm{i}} \simeq (1-\sigma_{\mathrm{c}})^{1/4}$. 
\end{itemize}
Further, if the helicity barrier both results in dissipation by ICWs and plays a role in the formation of the transition range, then the properties of the transition range would be expected to correlate with the presence of ICWs. Consistent with this, the transition range has been observed to be steeper for intervals with higher levels of circular polarisation \cite{2024NatAs.tmp...24B, 2024arXiv240610446B}.

In this paper, the dependencies of the location and slope of transition ranges of magnetic spectra, obtained using data from NASA's Parker Solar Probe (PSP) \cite{2016SSRv..204....7F, Raouafi_2023}, on solar wind parameters (cross helicity, plasma beta and degree of circular polarisation) are shown to be consistent with the presence of the helicity barrier. The values of some of these parameters necessary for the barrier to form are identified, and further are found to be common in solar wind. This suggests the helicity barrier to be an important consideration for understanding solar wind heating.

\section{Data and methods}
PSP data from orbits 1-10 were used. The magnetic field data used was the SCaM data product \cite{2020JGRA..12527813B}, which is produced by merging data from the search-coil magnetometer (SCM) and the fluxgate magnetometer (MAG) instruments of FIELDS instrument suite \cite{2016SSRv..204...49B}. From orbit 2 onwards data is not available for one axis of SCM so all trace magnetic spectra used were constructed using measurements from only two axes. Signatures of PSP's reaction wheels were identified and removed from the time series using a method similar to Shankarappa et al. \cite{2023ApJ...946...85S}. The ion velocity and temperature data were provided by the SPAN-I instrument of the SWEAP instrument suite \cite{2016SSRv..204..131K}. The velocity data were obtained through bi-Maxwellian fits \cite{2021A&A...650L...1W} where available, only being used where at least $3\phi$ bins were fitted to, with moments being used otherwise. Density data were obtained from quasi-thermal noise (QTN) measurements of the Radio Frequency Spectrometer Low Frequency Receiver  \cite{2020ApJS..246...44M}.

The data were split into non-overlapping 6 minute intervals. Intervals where more than 1\% of magnetic field data or 10\% of velocity data were missing were removed. The velocity distribution of the solar wind is not always sufficiently in the field of view of SPAN-I  \cite{2016SSRv..204..131K}, to account for this any interval where the SPAN-I measured density was less than 10\% of the QTN measured density were also discarded.

\begin{figure}[ht!]
\centering
\includegraphics[width=\columnwidth,trim=0 0 0 0,clip]{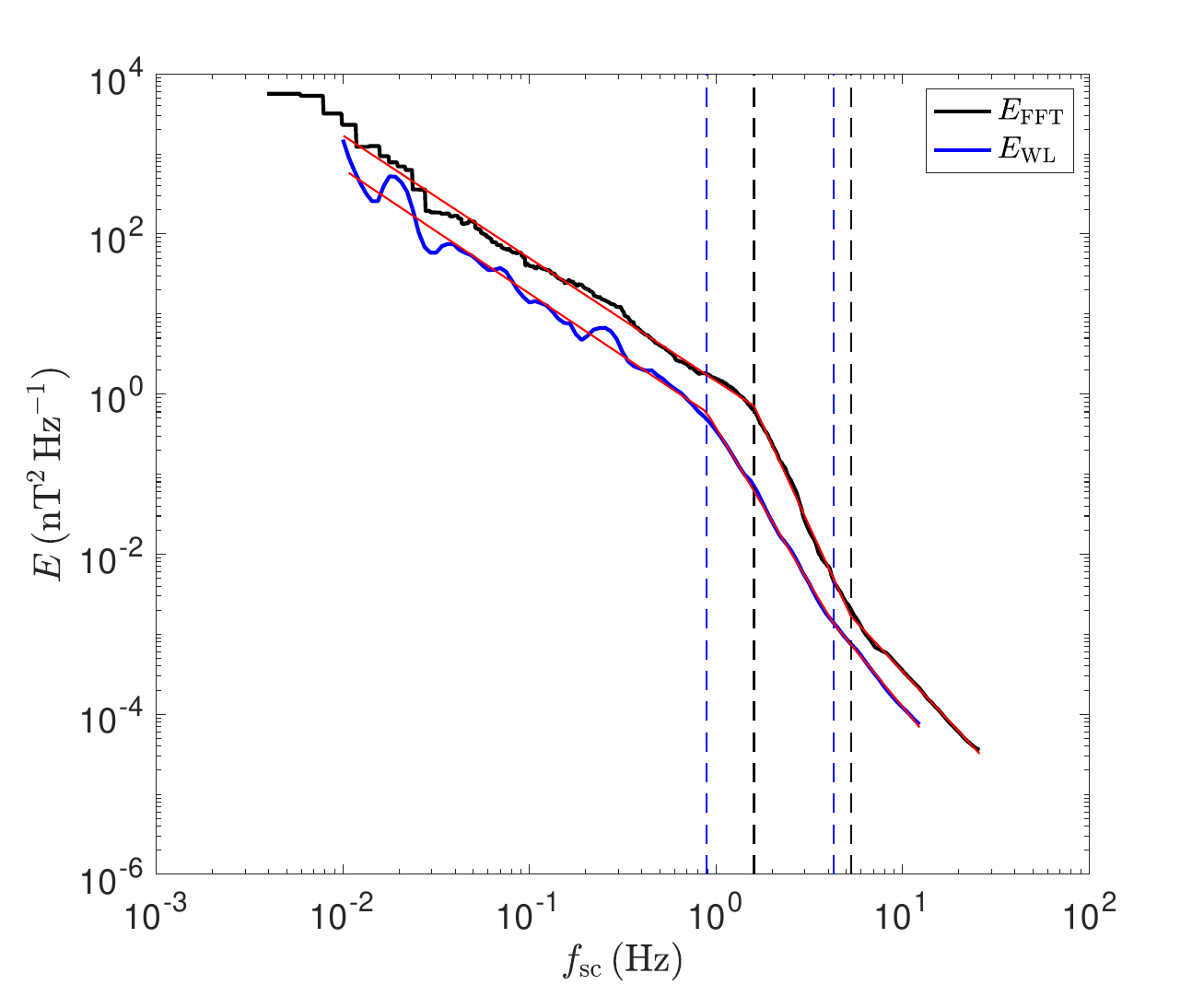}
\caption{Example magnetic spectra, $E$, obtained PSP data from 1 November 2018 20:30 to 20:36. The black line is obtained by FFT whereas the blue line has been obtained using a wavelet transform with power removed for frequencies and times where the degree of circular polarisation is high, as described in the text. The blue line has been artificially shifted to allow the spectra to be easily compared.  The dashed lines mark the break frequencies obtained from the three power law fits, and the red lines show those fits, for each spectrum.}
\label{fig:spectrum}
\end{figure}

A procedure similar to that of Shankarappa et al.\ \cite{2023ApJ...946...85S} was used to reduce the impact of ICW signatures on the spectra analysed. A time series of the degree of circular polarisation, 
\begin{equation} \label{eq:polarisation}
\sigma_{\mathrm{p}} = \frac{2 \, \mathrm{Im} (\tilde{B}_1 \tilde{B}_2^*)}{\tilde{B}_1^2 + \tilde{B}_2^2},
\end{equation}
where $\tilde{B}_1$ and $\tilde{B}_2$ are components of the magnetic field perpendicular to the local mean field in frequency space, was constructed for a discrete set of frequencies using wavelet transforms. $\sigma_{\mathrm{p}}$ has an absolute value of 1 for purely circularly polarised waves, because ICWs are circularly polarised its value can therefore be used to identity intervals where they may be present. The final spectra were obtained from wavelet transforms of the magnetic field time series but with times and frequencies where $|\sigma_{\mathrm{p}}|>0.7$ discarded \cite{2020ApJS..246...66B}. An example of spectra before and after this procedure was applied is shown in Figure \ref{fig:spectrum}. Note the clear bump around $2 \,\mathrm{Hz}$ in the pre-procedure black spectrum, which is not present in the post-procedure blue spectrum. If this bump were not removed it would result in the transition range being measured to be artificially steeper than its true value, thanks to the presence of ICWs.

After the reduction in the ICW signature was performed, a least squares regression procedure was used to fit each spectrum to a three power law function,
\begin{equation} \label{eq:3PL_fit}
E(f_\mathrm{sc}) \propto
   \begin{cases}
   f_\mathrm{sc}^{-\alpha_\mathrm{I}} & f_\mathrm{sc}<f_1 \\
   f_\mathrm{sc}^{-\alpha_\mathrm{T}} & f_1<f_\mathrm{sc}<f_2 \\
   f_\mathrm{sc}^{-\alpha_\mathrm{K}} & f_\mathrm{sc}>f_2 \\
   \end{cases}
\end{equation}
where $\alpha_\mathrm{I}$, $\alpha_\mathrm{T}$ and $\alpha_\mathrm{K}$ are the inertial, transition and kinetic spectral indices respectively. The fit has six parameters: the three indices, two break points and a $y$ offset. The obtained fit for the spectrum of Figure \ref{fig:spectrum} is shown in that figure. It is also shown for the spectrum before the ICW signature was removed, for comparison.

After the values of the fit parameters were obtained, intervals where $\alpha_\mathrm{K} > \alpha_\mathrm{T} + 0.1$ were discarded. This removed intervals where the fit fails, as we expect $\alpha_\mathrm{K} < \alpha_\mathrm{T}$, but the 0.1 tolerance prevents intervals being discarded if there is simply not a clear transition range and so $\alpha_\mathrm{T} \sim \alpha_\mathrm{K}$. After all the filters described in this section were applied 12334 intervals remained.

\section{Results}
\subsection{Dependence of transition range index on cross helicity and plasma beta}
If the helicity barrier mechanism is active in the solar wind it would be expected that a more prominent transition range would be observed in regions of high imbalance and low ion plasma beta, as these are conditions necessary for the barrier to exist. The dependencies of the transition range index on these parameters was therefore considered.

\begin{figure}[ht!]
\centering
\includegraphics[width=\linewidth,trim=0 0 0 0,clip]{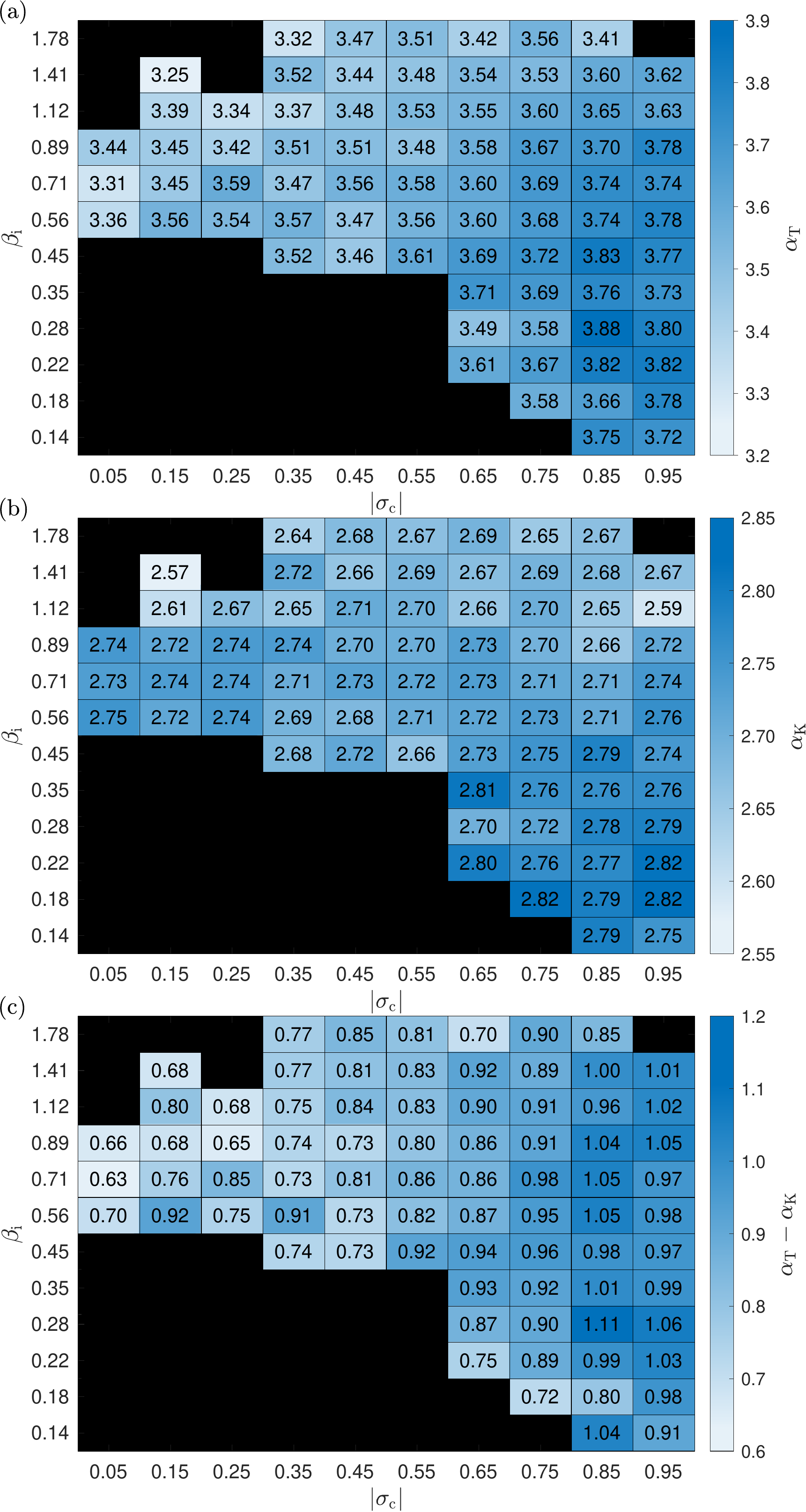}
\caption{(a), (b) and (c) show, respectively, the median values of $\alpha_\mathrm{T}$, $\alpha_\mathrm{K}$ and $\alpha_\mathrm{T}-\alpha_\mathrm{K}$ for bins of the intervals by absolute cross helicity, $|\sigma_{\mathrm{c}}|$, and ion plasma beta, $\beta_{\mathrm{i}}$. Black squares indicate that there were fewer than 30 intervals in that bin.}
\label{fig:beta_sigma_heatmaps}
\end{figure}

Figure \ref{fig:beta_sigma_heatmaps}(a) shows the median value of the transition range spectral index, $\alpha_\mathrm{T}$, for intervals binned by the normalised cross helicity,
\begin{equation} \label{eq:cross_helicity}
\sigma_{\mathrm{c}} = \frac{\langle \delta \boldsymbol{z}^{+2} - \delta \boldsymbol{z}^{-2} \rangle}{\langle \delta \boldsymbol{z}^{+2} + \delta \boldsymbol{z}^{-2} \rangle},
\end{equation}
and the ion plasma beta,
\begin{equation} \label{eq:beta}
\beta_{\mathrm{i}} = \frac{n_{0\mathrm{i}} k_\mathrm{B} T_{0\mathrm{i}}}{B_0^2/2\mu_0},
\end{equation}
where $n_{0\mathrm{i}}$ is the mean ion number density, $T_{0\mathrm{i}}$ the background ion temperature and $B_0$ the background field strength. The figure shows a clear trend of increasing $\alpha_\mathrm{T}$ with increasing $|\sigma_{\mathrm{c}}|$. There also appears to be a tendency for $\alpha_\mathrm{T}$ to increase for decreasing $\beta_{\mathrm{i}}$, at least until lower values of $\beta_{\mathrm{i}}$ are reached. These are both consistent with predictions of the helicity barrier.

Figure \ref{fig:beta_sigma_heatmaps}(b) shows the equivalent for $\alpha_\mathrm{K}$ and Figure \ref{fig:beta_sigma_heatmaps}(c) for $\alpha_\mathrm{T}-\alpha_\mathrm{K}$. The former shows no apparent trend of $\alpha_\mathrm{K}$ on $|\sigma_{\mathrm{c}}|$, in contrast to Bowen et al.\ \cite{2024NatAs.tmp...24B}. However, that work measured $\alpha_\mathrm{K}$ at a fixed scale of $k d_\mathrm{i} = 10$, considerably deeper into kinetic range than measured here. The latter shows a tendency for $\alpha_\mathrm{T}-\alpha_\mathrm{K}$ to increase with $|\sigma_{\mathrm{c}}|$, demonstrating that it is not just that the transition range becomes steeper with greater imbalance, but that there is a stronger break between it and the kinetic range. These features would be expected with the helicity barrier active.  

\begin{figure*}[ht!]
\centering
\includegraphics[width=1.27\columnwidth,trim=0 0 0 0,clip]{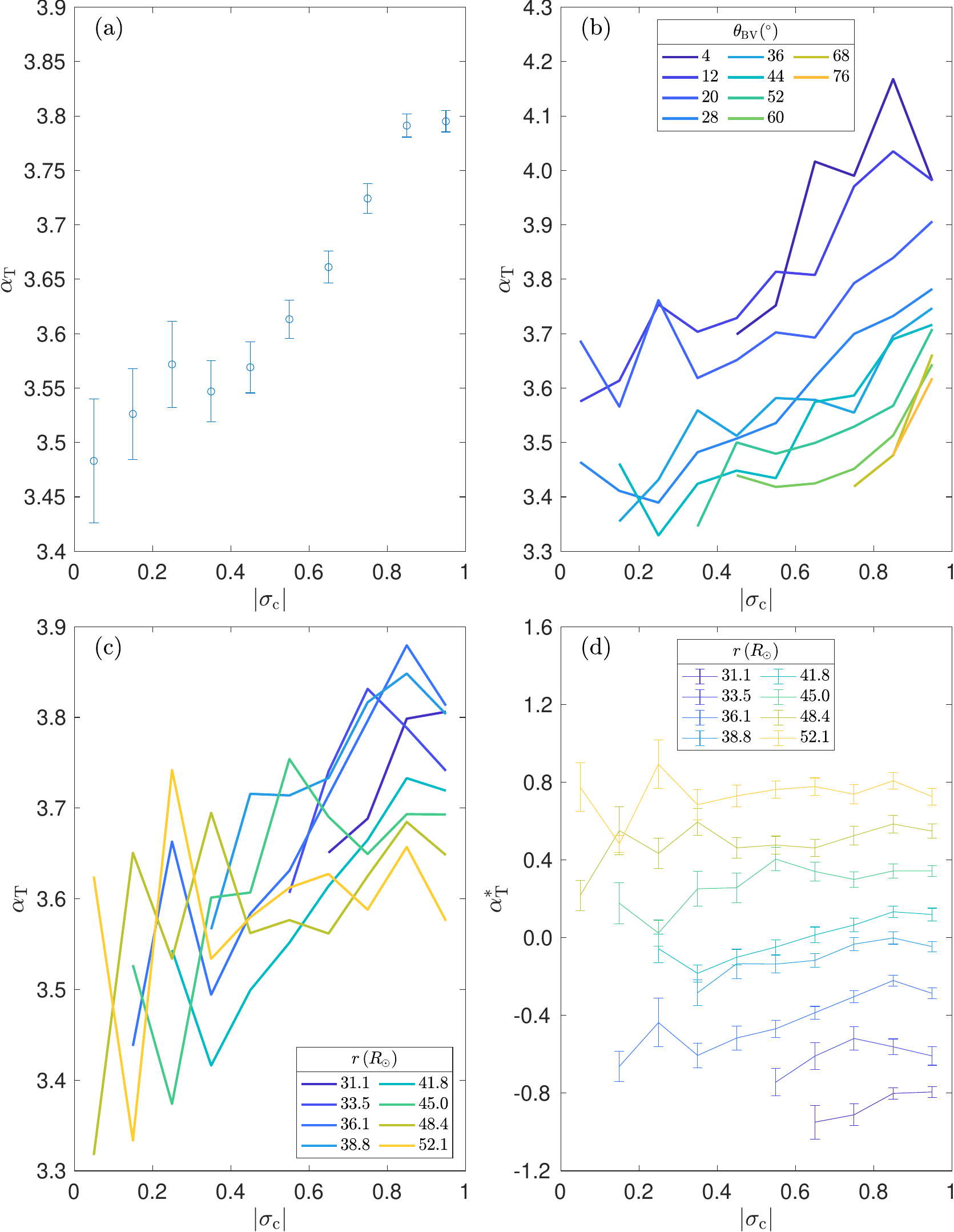}
\caption{(a) The mean value of transition range index, $\alpha_\mathrm{T}$, for the intervals binned by absolute cross helicity, $|\sigma_{\mathrm{c}}|$, with associated standard errors. (b) The mean value of $\alpha_\mathrm{T}$ against $|\sigma_{\mathrm{c}}|$ with each coloured line restricted to intervals in bins of sampling angle, $\theta_{\mathrm{BV}}$. (c) As for (b), but with the coloured lines corresponding to bins of radial distance, $r$. (d) The data of plot (c), but with lines artificially shifted to allow each to be seen. Standard errors are also shown.}
\label{fig:alpha_sigma}
\end{figure*}

The mean value of $\alpha_\mathrm{T}$ for intervals binned by $|\sigma_{\mathrm{c}}|$, with associated standard errors, is shown in Figure \ref{fig:alpha_sigma}(a), again demonstrating the trend discussed above. The transition range clearly steepens with imbalance for $|\sigma_{\mathrm{c}}| \gtrsim 0.4$. To confirm that this trend is robust it was verified that the dependence holds when further parameters (on which $\alpha_\mathrm{T}$ could plausibly depend on) are held approximately constant. The first parameter considered was $\theta_{\mathrm{BV}}$, the angle between the mean magnetic field and mean solar wind velocity for each interval, as measured in the spacecraft frame. This sampling angle is important to consider as solar wind turbulence is known to be anisotropic \cite{2012SSRv..172..325H, chen_2016}, with Duan et al.\ \cite{2021ApJ...915L...8D} measuring the transition index to be steeper for spacecraft trajectories parallel to the background field. The result of this test is shown in Figure \ref{fig:alpha_sigma}(b), which again shows the mean value of $\alpha_\mathrm{T}$ for intervals binned by $|\sigma_{\mathrm{c}}|$, but with each coloured line calculated only from intervals within a set range of $\theta_{\mathrm{BV}}$. As the lines vary from yellow to blue $\theta_{\mathrm{BV}}$ becomes increasingly parallel. Consistent with Duan et al.\ \cite{2021ApJ...915L...8D}, parallel sampled intervals do indeed have steeper transition ranges than perpendicularly sampled intervals but, across bins, the trend with the imbalance remains robust.

The second factor considered was the radial distance from the Sun, $r$. As cross helicity is known to vary with $r$ \cite{1987JGR....9211021R, 1995SSRv...73....1T, 1998JGR...103.6521B, 2000JGR...10515959B, 2004GeoRL..3112803M, 2005GeoRL..32.6103B, 2007AnGeo..25.1913B, 2020ApJS..246...53C} any apparent trend with $|\sigma_{\mathrm{c}}|$ could actually result from other solar wind conditions changing with distance from the Sun. Figure \ref{fig:alpha_sigma}(c) shows the mean $\alpha_\mathrm{T}$ against $|\sigma_{\mathrm{c}}|$, with each coloured line determined using intervals in a set range of $r$ values. The dark blue line corresponds to the intervals closest to the Sun and yellow to the furthest. The same data is shown in Figure \ref{fig:alpha_sigma}(d) but with the lines artificially spaced vertically to enable them to be seen more clearly, and with standard errors provided for each data point. The trend with $|\sigma_{\mathrm{c}}|$ clearly remains for most distance bins, particularly for those of lower $r$. The trend is less clear for intervals further from the Sun, possibly as plasma beta is generally higher for those intervals. From both these described tests, the observed trend on imbalance appears robust.

\begin{figure*}[ht!]
\centering
\includegraphics[width=1.27\columnwidth,trim=0 0 0 0,clip]{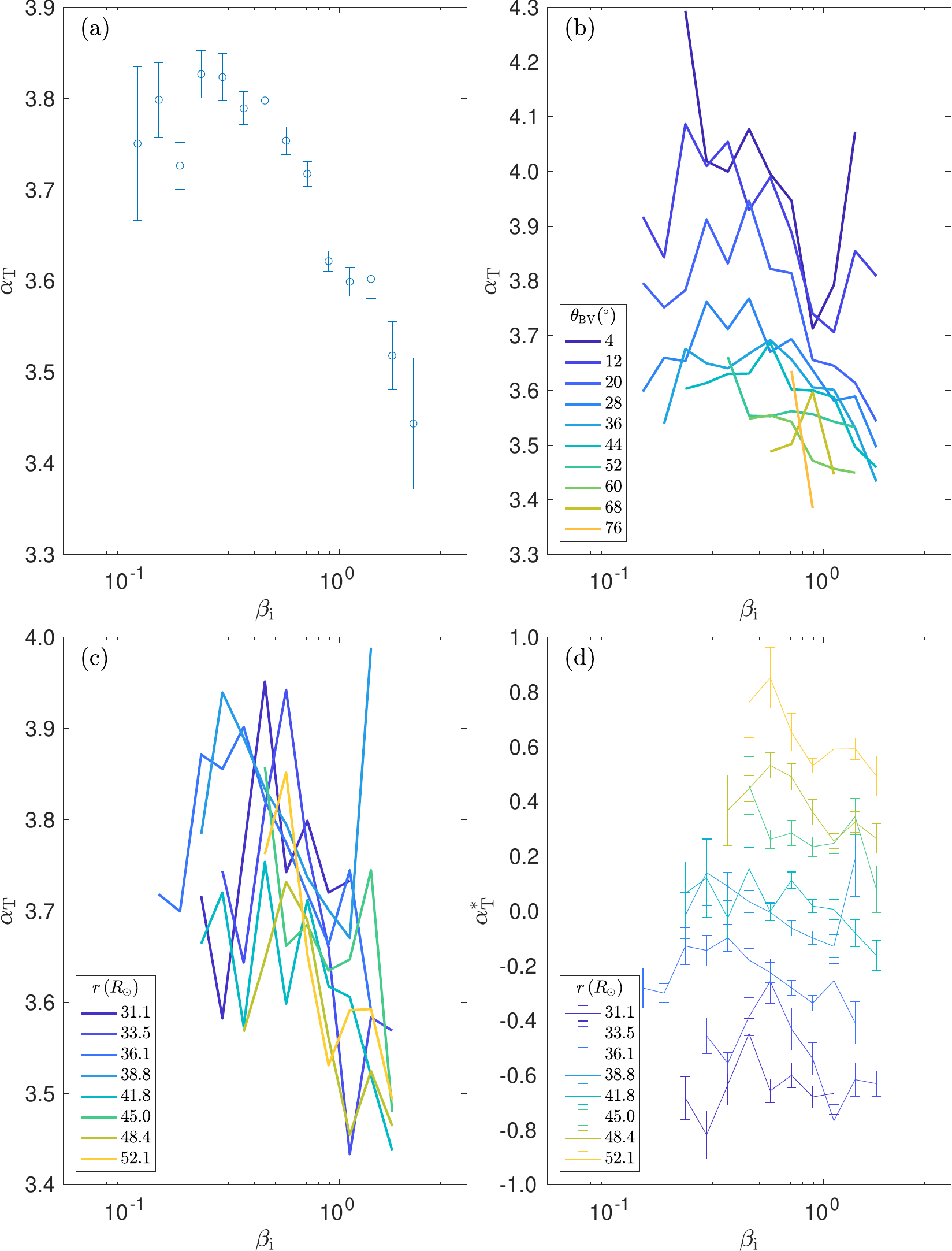}
\caption{(a) The mean value of transition range index, $\alpha_\mathrm{T}$, for the intervals binned by ion plasma beta, $\beta_{\mathrm{i}}$. (b) The mean value of $\alpha_\mathrm{T}$ against $\beta_{\mathrm{i}}$ with each coloured line restricted to intervals in bins of sampling angle, $\theta_{\mathrm{BV}}$. (c) As for (b), but with the coloured lines corresponding to bins of radial distance, $r$. (d) The data of plot (c), but with lines artificially shifted to allow each to be seen.}
\label{fig:alpha_beta}
\end{figure*}

The dependence of $\alpha_\mathrm{T}$ on $\beta_{\mathrm{i}}$ in Figure \ref{fig:beta_sigma_heatmaps}(a) can be seen more clearly in Figure \ref{fig:alpha_beta}(a), which shows the mean value of $\alpha_\mathrm{T}$ for intervals binned by $\beta_{\mathrm{i}}$. The transition range becomes steeper with decreasing $\beta_{\mathrm{i}}$ until saturating around $\beta_{\mathrm{i}}\simeq0.5$.   This is consistent with predictions of the helicity barrier, as it would not necessarily be expected that $\alpha_\mathrm{T}$ should continue to increase for ever decreasing $\beta_{\mathrm{i}}$, it only needs to be low enough for FLR-MHD to be a valid description of the plasma. As with the imbalance trends, it was verified that this apparent behaviour remains when the intervals are binned by either sampling angle or radial distance. Figure \ref{fig:alpha_beta}(b) is analogous to Figure \ref{fig:alpha_sigma}(b), showing the mean $\alpha_\mathrm{T}$ against $\beta_{\mathrm{i}}$, with each coloured line corresponding to a different $\theta_{\mathrm{BV}}$ bin. Once again, parallel measured intervals tend to have steeper transition ranges but the trend with $\beta_{\mathrm{i}}$ remains.  Figures \ref{fig:alpha_beta}(c) and \ref{fig:alpha_beta}(d) are analogous to Figures \ref{fig:alpha_sigma}(c) and \ref{fig:alpha_sigma}(d), respectively. Both plots show the mean value of $\alpha_\mathrm{T}$ against $\beta_{\mathrm{i}}$ for coloured lines corresponding to radial distance bins, with the lines artificially spaced in Figure \ref{fig:alpha_beta}(d). The observed behaviour of $\alpha_\mathrm{T}$ with $\beta_{\mathrm{i}}$ appears robust to this test also, with the same trend still apparent. 

\subsection{Dependence of transition index on degree of circular polarisation}
A further parameter considered was the degree of left-handed circular polarisation of the intervals, $\sigma_{\mathrm{p}}^{\mathrm{LH}}$. If the helicity barrier results in both  the generation of ICWs, which are LH circularly polarised, and a more prominent transition range, then we should expect $\alpha_\mathrm{T}$ to be positivity correlated with $\sigma_{\mathrm{p}}^{\mathrm{LH}}$.

For each interval $\sigma_{\mathrm{p}}$ (Eq \eqref{eq:polarisation}) was calculated as a function of frequency using Fast Fourier Transforms of the magnetic field components perpendicular to the mean field of the interval. The values of $\sigma_{\mathrm{p}}$ with a sign corresponding to left-handed polarisation were then extracted and the resulting spectrum smoothed by averaging over a sliding window. This produced a measure of the degree of left-handed circular polarisation against frequency. $\sigma_{\mathrm{p}}^{\mathrm{LH}}$ was defined to be the maximum value of this function for frequencies greater than half the low frequency break point, $f_1$. While the Doppler shift can cause the polarisation measured in the spacecraft frame to invert from that of the plasma frame for waves with $\boldsymbol{k}\cdot \boldsymbol{V}_\mathrm{sw}<0$ \cite{2020ApJS..246...66B}, the following results are not changed significantly if the absolute polarisation is used rather than the left-hand polarisation.

\begin{figure*}[ht!]
\centering
\includegraphics[width=\linewidth,trim=0 0 0 0,clip]{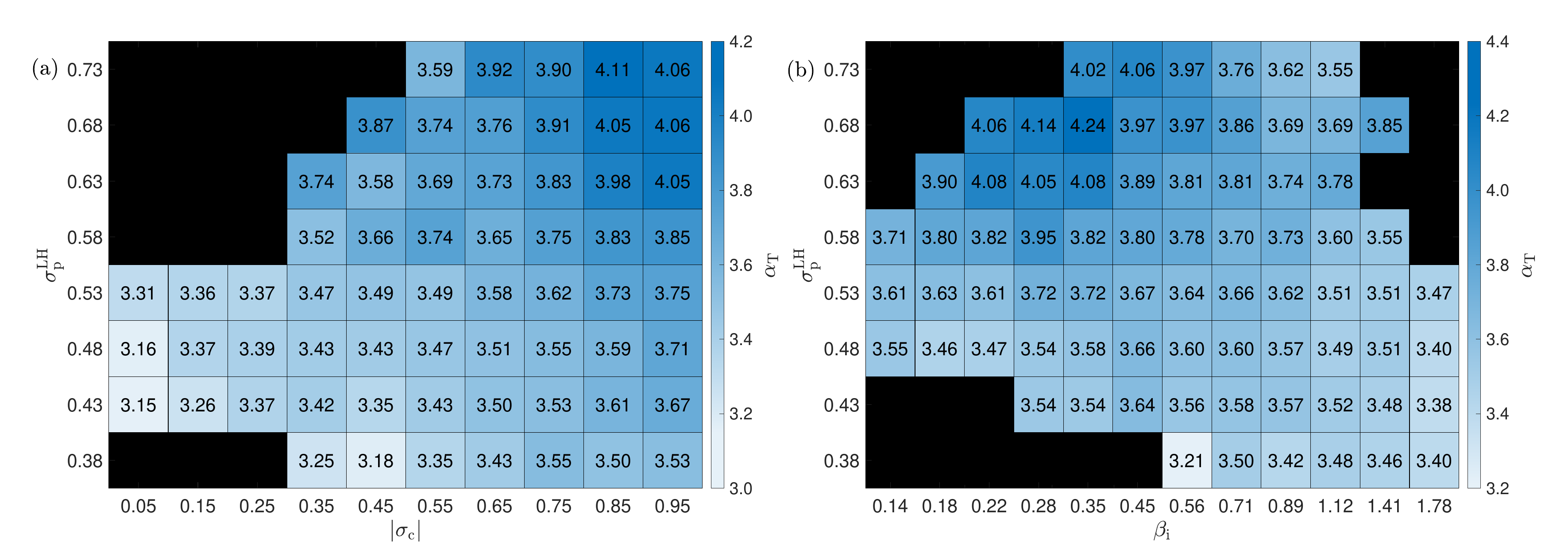}
\caption{(a) The median value of the transition range index, $\alpha_\mathrm{T}$,  for the intervals binned by the degree of circular polarisation, $\sigma_{\mathrm{p}}^{\mathrm{LH}}$, and the absolute cross helicity, $|\sigma_{\mathrm{c}}|$. (b) The median value of $\alpha_\mathrm{T}$,  for the intervals binned by  $\sigma_{\mathrm{p}}^{\mathrm{LH}}$ and ion plasma beta, $\beta_{\mathrm{i}}$. For both black squares indicate that there were fewer than 30 intervals in that bin.}
\label{fig:lh_heatmaps}
\end{figure*}

Figure \ref{fig:lh_heatmaps}(a) shows the median value of $\alpha_\mathrm{T}$ for intervals binned by $\sigma_{\mathrm{p}}^{\mathrm{LH}}$ and $|\sigma_{\mathrm{c}}|$, while Figure \ref{fig:lh_heatmaps}(b) shows this for intervals binned by $\sigma_{\mathrm{p}}^{\mathrm{LH}}$ and $\beta_{\mathrm{i}}$ (note that $\sigma_{\mathrm{p}}^{\mathrm{LH}}$ is calculated from the magnetic time series without ICW signatures being removed and so values of greater than 0.7 are possible). There is a clear positive correlation between $\sigma_{\mathrm{p}}^{\mathrm{LH}}$ and $\alpha_\mathrm{T}$, consistent with predictions of the helicity barrier. This correlation holds for both approximately constant $|\sigma_{\mathrm{c}}|$ or $\beta_{\mathrm{i}}$.

\begin{figure*}[ht!]
\centering
\includegraphics[width=1.27\columnwidth,trim=0 0 0 0,clip]{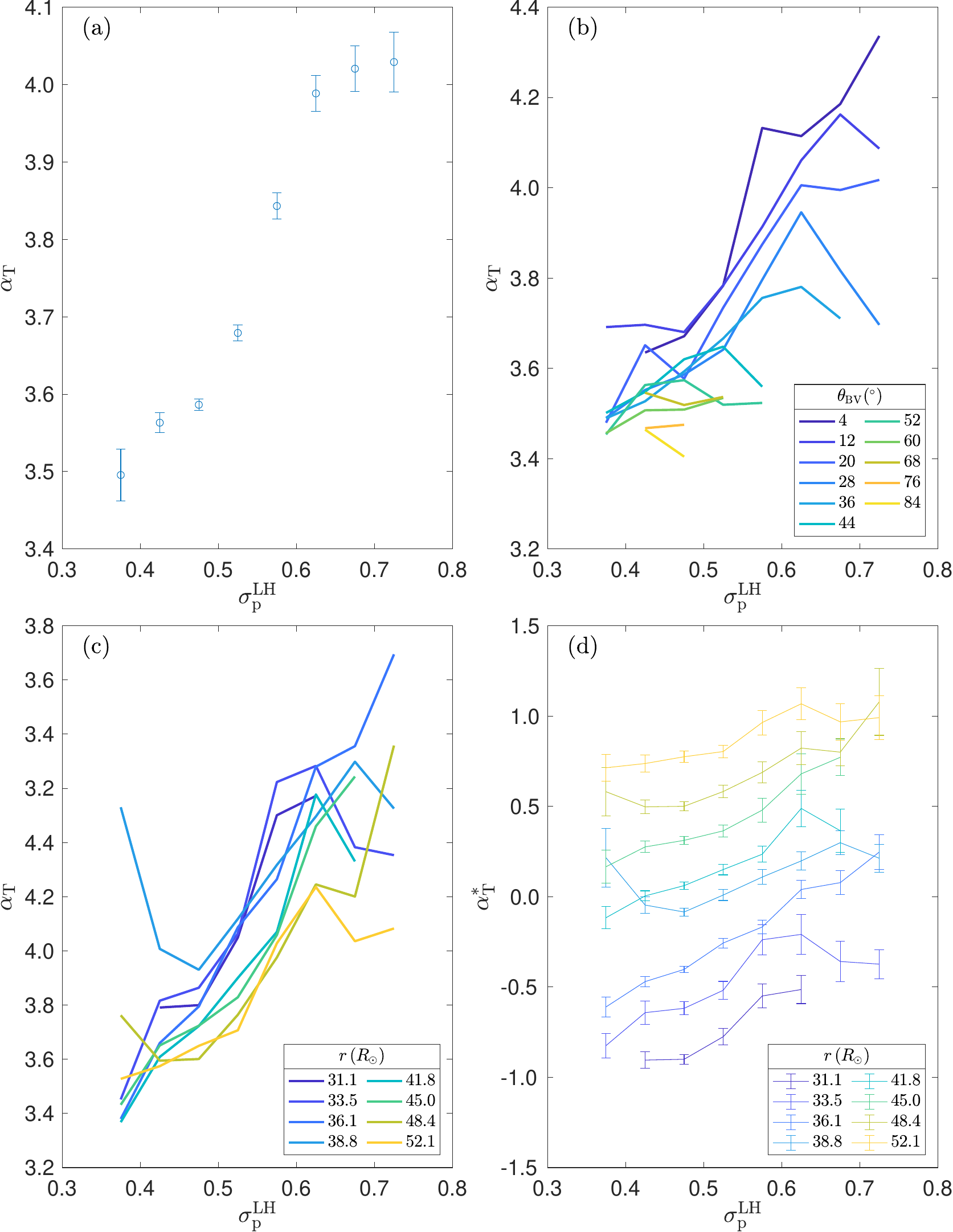}
\caption{(a) The mean value of transition range index, $\alpha_\mathrm{T}$, for the intervals binned by the degree of left-handed circular polarisation, $\sigma_{\mathrm{p}}^{\mathrm{LH}}$. (b) The mean value of $\alpha_\mathrm{T}$ against $\sigma_{\mathrm{p}}^{\mathrm{LH}}$ with each coloured line restricted to intervals in bins of sampling angle, $\theta_{\mathrm{BV}}$. (c) As for (b), but with the coloured lines corresponding to bins of radial distance, $r$. (d) The data of plot (c), but with lines artificially shifted to allow each to be seen.}
\label{fig:alpha_lh}
\end{figure*}

This trend can again be seen in Figure \ref{fig:alpha_lh}(a), which shows the mean value of $\alpha_\mathrm{T}$ for intervals binned by $\sigma_{\mathrm{p}}^{\mathrm{LH}}$, with associated standard errors. It was verified that the behaviour holds when the sampling angle or radial distance was held approximately constant, as was done for the trends found with $|\sigma_{\mathrm{c}}|$ and $\beta_{\mathrm{i}}$. Figure \ref{fig:alpha_lh}(b) shows this test for $\theta_{\mathrm{BV}}$, with each coloured line of mean $\alpha_\mathrm{T}$ against $\sigma_{\mathrm{p}}^{\mathrm{LH}}$ corresponding to intervals from a sampling angle bin. While intervals with low $\theta_{\mathrm{BV}}$ tend to have higher $\alpha_\mathrm{T}$, as we also show in Figures \ref{fig:alpha_sigma}(b) and \ref{fig:alpha_beta}(b), the trend with $\sigma_{\mathrm{p}}^{\mathrm{LH}}$ remains. Figures \ref{fig:alpha_lh}(c) and (d) show the equivalent test for radial distance, with each coloured line corresponding to a $r$ bin. From these it is clear that the positive $\alpha_\mathrm{T}$ trend with $\sigma_{\mathrm{p}}^{\mathrm{LH}}$ also remains when distance is held approximately constant. The trend therefore appears robust.

\subsection{Dependencies of the break points}
\subsubsection{High frequency break point}
The behaviour of the transition range high frequency break point, $f_2$, with varying solar wind parameters was also considered. These frequencies were converted to wavenumbers, $k_2$, through application of the Taylor hypothesis \cite{1938RSPSA.164..476T}, and normalised to the ion gyroradius, $\rho_{\mathrm{i}}$. The sampling angle, $\theta_{\mathrm{BV}}$, was used to obtain the perpendicular wavenumber through $k_{\perp,2} =k_{2} \sin{\theta_{\mathrm{BV}}}$. Unlike with $\alpha_\mathrm{T}$, the helicity barrier does not make clear predictions for the behaviour of $k_{\perp 2}$.

\begin{figure}[ht!]
\centering
\includegraphics[width=\columnwidth,trim=0 0 0 0,clip]{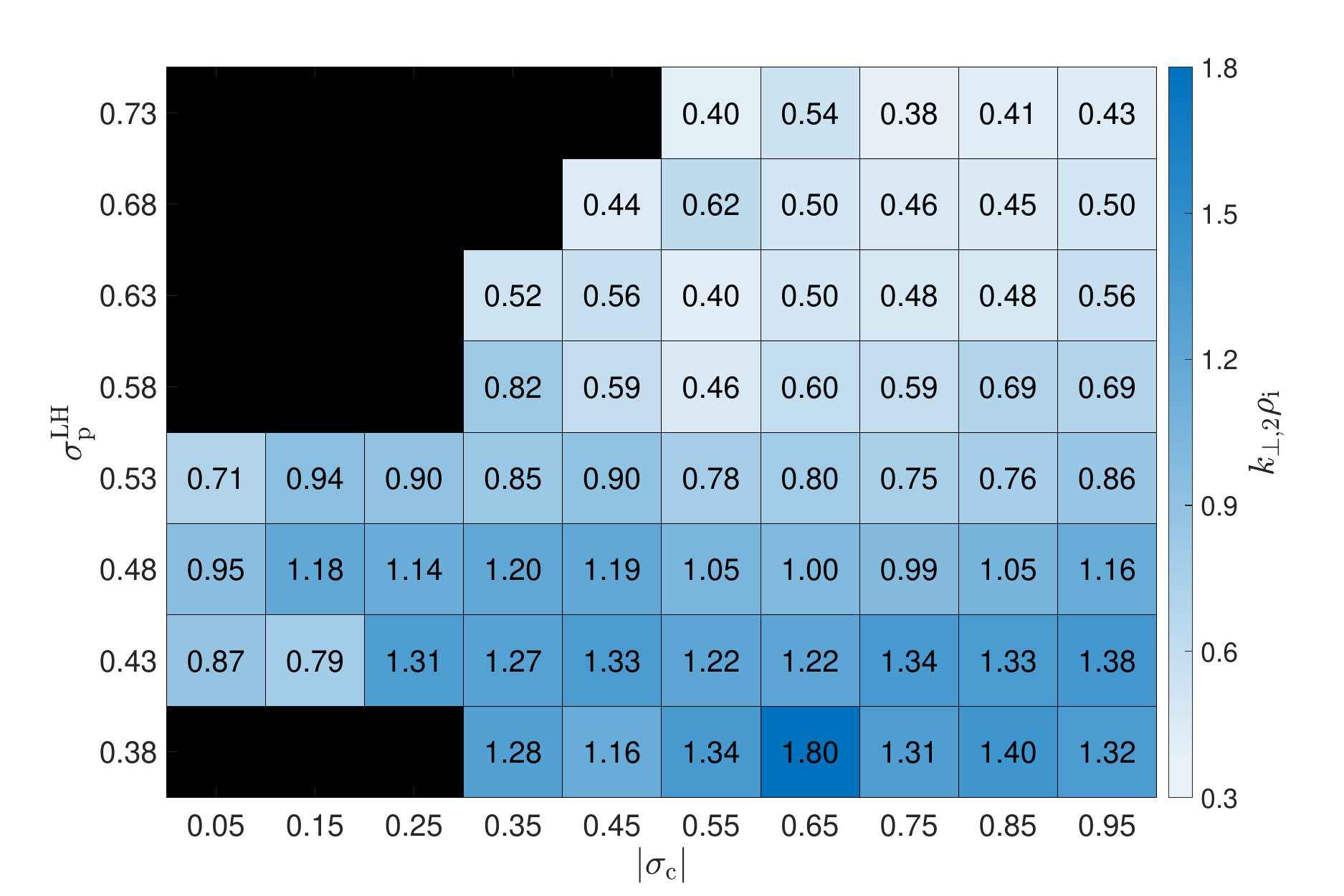}
\caption{The median value of the high frequency break point, $k_{\perp,2} \rho_{\mathrm{i}}$, for intervals binned by the normalised cross helicity, $|\sigma_{\mathrm{c}}|$, and the degree of left-handed polarisation, $\sigma_{\mathrm{p}}^{\mathrm{LH}}$. Black bins have fewer than 30 intervals within them.}
\label{fig:k2_heatmap}
\end{figure}

\begin{figure*}[ht!]
\centering
\includegraphics[width=1.27\columnwidth,trim=0 0 0 0,clip]{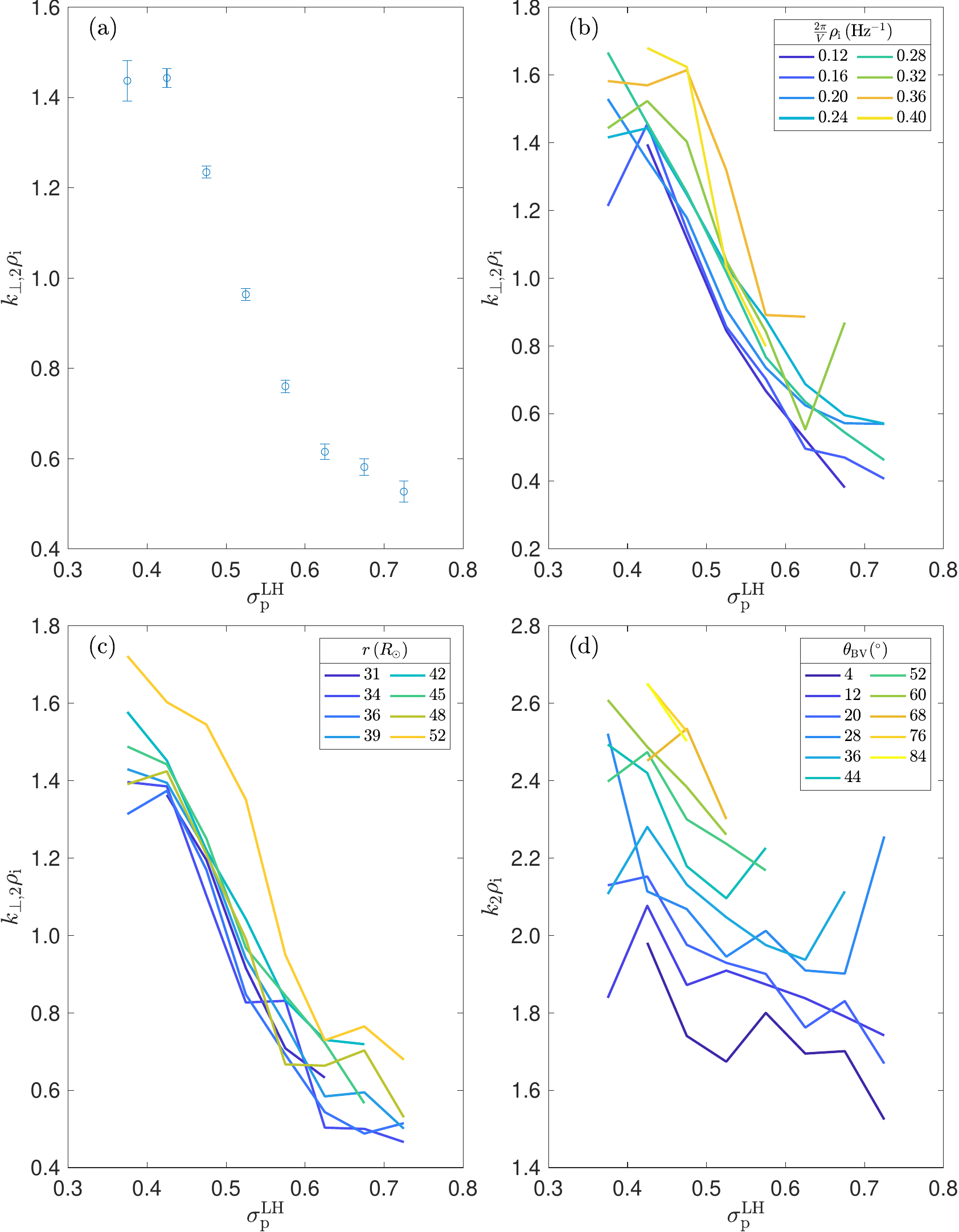}
\caption{(a) The mean value of high frequency break point, $k_{\perp,2} \rho_{\mathrm{i}}$, for the intervals binned by the degree of left-handed circular polarisation, $\sigma_{\mathrm{p}}^{\mathrm{LH}}$. (b) The mean value of $k_{\perp,2} \rho_{\mathrm{i}}$ against $\sigma_{\mathrm{p}}^{\mathrm{LH}}$, with each coloured line restricted to intervals in bins of the conversion factor from frequencies to $k_2 \rho_{\mathrm{i}}$, $\frac{2\pi}{V}\rho_{\mathrm{i}}$. (c) As for (b), but with the coloured lines corresponding to bins of radial distance, $r$. (d) The mean value of $k_2 \rho_{\mathrm{i}}$ against $\sigma_{\mathrm{p}}^{\mathrm{LH}}$ with each coloured line restricted to intervals in bins of sampling angle, $\theta_{\mathrm{BV}}$.}
\label{fig:k2_lh}
\end{figure*}

Figure \ref{fig:k2_heatmap} shows the median value of $k_{\perp,2} \rho_{\mathrm{i}}$ for intervals binned by $|\sigma_{\mathrm{c}}|$ and $\sigma_{\mathrm{p}}^{\mathrm{LH}}$. There is a clear negative correlation of $k_{\perp,2} \rho_{\mathrm{i}}$ with $\sigma_{\mathrm{p}}^{\mathrm{LH}}$. This is also clear in Figure \ref{fig:k2_lh}(a), which shows the mean value of $k_{\perp,2} \rho_{\mathrm{i}}$ for intervals binned by $\sigma_{\mathrm{p}}^{\mathrm{LH}}$. There is no clear trend with $|\sigma_{\mathrm{c}}|$ across the $\sigma_{\mathrm{p}}^{\mathrm{LH}}$ bins. The behaviour with $\beta_{\mathrm{i}}$ was not considered as $\beta_{\mathrm{i}}$ and $\rho_{\mathrm{i}}$ are proportional to $T_{0\mathrm{i}}/B_0^2$ and $\sqrt{T_{0\mathrm{i}}}/B_0$ respectively, making it difficult to untangle any $k_2 \rho_{\mathrm{i}}$ trend on $\beta_{\mathrm{i}}$ from a mutual scaling with these solar wind parameters. 

As in previous sections, it was verified that the trend of $k_{\perp,2} \rho_{\mathrm{i}}$ on $\sigma_{\mathrm{p}}^{\mathrm{LH}}$ is not the result of a shared dependency on another parameter. The factor used to convert $f_1$ to $k_2 \rho_{\mathrm{i}}$, $\frac{2\pi}{V}\rho_{\mathrm{i}}$, where $V$ is the solar wind speed, was one such parameter considered. If the position of the break depends on $\sigma_{\mathrm{p}}^{\mathrm{LH}}$ in a physically meaningful way then the trend should remain if this conversion factor is held constant. Figure \ref{fig:k2_lh}(b) shows the result of this test, with each coloured line corresponding to a plot of  $k_{\perp,2} \rho_{\mathrm{i}}$ against $\sigma_{\mathrm{p}}^{\mathrm{LH}}$ for a $\frac{2\pi}{V}\rho_{\mathrm{i}}$ bin. The trend with $\sigma_{\mathrm{p}}^{\mathrm{LH}}$ can still be seen for each bin and so it is robust with respect to this test. 

It was further verified that the trend with $\sigma_{\mathrm{p}}^{\mathrm{LH}}$ holds independent of radial distance, checked by the same method as in previous sections, with the results shown in Figure \ref{fig:k2_lh}(c). It is clear that when distance is held approximately constant, the trend remains. 

Figure \ref{fig:k2_lh}(d) demonstrates that the trend also remains independent of sampling angle, which is important to confirm given the dependence of $\sigma_{\mathrm{p}}^{\mathrm{LH}}$ on $\theta_{\mathrm{BV}}$. Here $k_2 \rho_{\mathrm{i}}$ was considered, rather than $k_{\perp,2} \rho_{\mathrm{i}}$, to further isolate the trend from any $\theta_{\mathrm{BV}}$ effects. 

Intervals with high $\sigma_{\mathrm{p}}^{\mathrm{LH}}$ should have spectra that are more significantly affected by the procedure used to remove ICW signatures than those with low $\sigma_{\mathrm{p}}^{\mathrm{LH}}$, as the procedure relies on $\sigma_{\mathrm{p}}$ to remove the signatures. A possible effect of removing the ICW signatures is the introduction of a small kink to the spectrum at the signature's high frequency edge, which could be mistaken for $f_2$. This could introduce a $\sigma_{\mathrm{p}}^{\mathrm{LH}}$ dependent bias to the $k_{\perp,2} \rho_{\mathrm{i}}$ measurements. To ensure that the found trend of $k_{\perp,2} \rho_{\mathrm{i}}$ on $\sigma_{\mathrm{p}}^{\mathrm{LH}}$ is not a result of this, a procedure was developed to find the high frequency end of the ICW signature, $f^*$, so that intervals where it is close to $f_2$ could be excluded. Intervals where $f^*/1.2 < f_2 < 1.2 f^*$ were discarded and the above analysis repeated, exhibiting no significant change in any of the results. The trend is therefore not the result of this potential complication.

\subsubsection{Low frequency break point}
The behaviour of $k_{\perp,1} \rho_{\mathrm{i}}$ was also considered. The presence of ICW signatures can make it unclear as to where the low frequency break should be placed, with $f_1$ often being determined to be at frequencies a little higher than the peak of the ICW bump (as this is where the spectrum is particularly steep) before the procedure to remove the signatures is applied, and at the low frequency edge of the bump after the signatures have been removed (as this can introduce a kink to the spectrum). As a result, only intervals with $\sigma_{\mathrm{p}}^{\mathrm{LH}}<0.5$ were used when considering the behaviour of $k_{\perp,1} \rho_{\mathrm{i}}$. For such intervals the break locations identified agree sufficiently well before and after the signatures were removed to give confidence in the results obtained. Indeed, there are no significant differences in the results obtained for the two sets of $f_1$.  

A potential trend of decreasing $k_{\perp,1} \rho_{\mathrm{i}}$ for increasing $|\sigma_{\mathrm{c}}|$ was found, which would be consistent with the barrier's prediction. However, this trend was found not to hold when the conversion factor from $f_1$ to $k_1\rho_{\mathrm{i}}$ was held approximately constant and so we cannot report this possible trend with confidence. As the predicted $k_{\perp,1} \rho_{\mathrm{i}} \simeq (1-\sigma_{\mathrm{c}})^{1/4}$ trend is rather weak it is perhaps unsurprising that it could not be untangled from dependencies on other solar wind parameters. Given the constraint on the value of $\sigma_{\mathrm{p}}^{\mathrm{LH}}$ of the intervals used, a trend with the degree of circular polarisation was not considered.

\section{Discussion}
The results of this paper suggest the helicity barrier mechanism plays a significant role in the heating of the solar wind. We have shown that properties of the turbulence transition range vary with solar wind conditions in a manner consistent with the presence of the helicity barrier and have determined critical values of parameters necessary for the barrier to form. As the values of these parameters are common in the inner heliosphere the barrier should be frequently active in the solar wind, thus resulting in the nature of its heating being dependent on solar wind conditions in a way that can help explain its observed properties.

In order for the barrier to form $\beta_{\mathrm{i}}$ must be sufficiently low for FLR-MHD to be valid. We have shown, for the first time, that the spectral index of the transition range, $\alpha_\mathrm{T}$, increases with decreasing ion plasma beta, $\beta_{\mathrm{i}}$, until reaching a plateau at around $\beta_{\mathrm{i}}\simeq0.5$. This indicates that a value of  $\beta_{\mathrm{i}}\simeq0.5$ is sufficient for the mechanism to be fully active, a value commonly observed in the solar wind, particularly for regions close to the Sun. 

It was also shown that intervals with high imbalance have steeper transition ranges than intervals of low imbalance. This result is consistent with previous work \cite{Huang_2021,Zhao_2022,2024NatAs.tmp...24B}, which we build on by demonstrating that it is not just that that the transition range is steeper for regions of high $|\sigma_{\mathrm{c}}|$, but that there is a stronger break with the kinetic range for regions of high $|\sigma_{\mathrm{c}}|$, a picture consistent with the presence of the barrier. Unlike previous work, we made use of a procedure to reduce ICW signatures in the magnetic spectra analysed and so were able to verify that the turbulent spectrum itself is steepening with increasing imbalance, rather than the trend being the result of bumps introduced on top of the spectrum by ICWs, the presence of which is correlated with $|\sigma_{\mathrm{c}}|$. We identified $|\sigma_{\mathrm{c}}| \simeq0.4$ as a critical value for the barrier, with $\alpha_\mathrm{T}$ being shown to clearly be increasing with imbalance for values greater than this. Once again this a condition regularly met for intervals close to the Sun.

It was further found that $\alpha_\mathrm{T}$ correlates positively with the degree of left-handed circular polarisation, $\sigma_{\mathrm{p}}^{\mathrm{LH}}$, consistent with the transition range and the generation of ICWs both being possible results of the helicity barrier. This is consistent with previous work \cite{2024NatAs.tmp...24B, 2024arXiv240610446B}, but again we were able to verify that the spectrum itself is steepening by reducing ICW signatures.

The behaviour of the high frequency transition range break point, $k_{\perp,2} \rho_{\mathrm{i}}$, was also considered and found to robustly decrease with increasing $\sigma_{\mathrm{p}}^{\mathrm{LH}}$. Such behaviour may have been expected for the low frequency break point, as its predicted trend with imbalance suggests it may move to larger scales when the barrier is active, but no such behaviour is predicted for  $k_{\perp,2} \rho_{\mathrm{i}}$. A potential explanation for this is to consider $\sigma_{\mathrm{p}}^{\mathrm{LH}}$ as a proxy for dissipation through ion-cyclotron resonance, such that higher values of $\sigma_{\mathrm{p}}^{\mathrm{LH}}$ indicate a greater rate of dissipation. The dominant Elsasser field in high  $\sigma_{\mathrm{p}}^{\mathrm{LH}}$ intervals may therefore dissipate to reach the power of the other Elsasser field at larger scales than for low $\sigma_{\mathrm{p}}^{\mathrm{LH}}$ intervals of similar imbalance. The transition range would hence end at a lower frequency, consistent with the observed trend.

Overall, a high degree of consistency between the helicity barrier and the behaviour of the magnetic spectrum in the solar wind was found, increasing confidence in the barrier's presence there. The rather modest critical values found of the parameters considered, $\beta_{\mathrm{i}}\lesssim0.5$ and $|\sigma_{\mathrm{c}}| \gtrsim 0.4$, suggests the barrier is frequently active in the solar wind, making it an important consideration in understanding its heating. 

By causing greater amplitudes of the turbulent cascade to be achieved at ion scales than standard turbulence theory would predict, the barrier may enable ion-cyclotron heating to take place, resulting in preferential heating of ions over electrons. This provides an explanation of ions being observed to be hotter than electrons in the solar wind. As the helicity barrier requires imbalanced turbulence to be active the degree to which these processes would be expected to take place would depend on the cross helicity, which is known to be correlated with the wind speed. The barrier therefore also provides an explanation for the observed variation in the ratio of ion to electron temperature between fast and slow wind streams. Thus the dependencies introduced by the helicity barrier, by diverging from the standard turbulence picture of an unimpeded cascade to dissipation scales, have significant physical consequences. These consequences have with the potential to help answer key open questions about the dynamics of the solar wind, this work supports the helicity barrier mechanism playing such a role.

\medskip
J.R.M. is supported by STFC studentship grant ST/V506989/1. C.H.K.C. and P.A.S. are supported by UKRI Future Leaders Fellowship MR/W007657/1. C.H.K.C. is supported by STFC Consolidated Grants ST/T00018X/1 and ST/X000974/1. J.S. and R.M. acknowledge the support of the Royal Society Te Ap\=arangi, through Marsden-Fund grant MFP-UOO2221 (J.S.) and MFP-U0020 (R.M.), as well as through the Rutherford Discovery Fellowship RDF-U001004 (J.S.). J.R.M. acknowledges support from the Perren Exchange Programme. We thank Lloyd Woodham for providing the SPAN fits dataset and Trevor Bowen for helpful discussions. PSP data are available at the SPDF (https://spdf.gsfc.nasa.gov).

\bibliography{McIntyre24}{}

\end{document}